\documentclass[12pt]{article}
\usepackage{graphicx}
\usepackage{amsmath}

\newcommand{\NN}{\hbox{I\kern-.2em\hbox{N}}}  
\newcommand{\ZZ}{{{\rm Z}\kern-.28em{\rm Z}}} 
\newcommand{\RR}{\mathop{{\rm I}\kern-.2em{\rm R}}\nolimits} 
\newcommand{\QQ}{\hbox{l\kern-.36em\hbox{Q}}}  
\newcommand{\CC}{\hbox{I\kern-.58em\hbox{C}}}

\begin{document}
\title{Simultaneous elements of reality for incompatible properties by exploiting locality}
\author{Angela Sestito\footnote{The present work is supported by the European Commission,
European Social Fund and  by the Calabria Region, Regional Operative
Program (ROP)
Calabria ESF 2007/2013 - IV Axis Human Capital - Operative Objective M2- Action D.5}\\
{\small Dipartimento
di Matematica, Universit\`a della Calabria, Italy}\\
{\small and}\\
{\small INFN -- gruppo collegato di Cosenza, Italy}\\
{\small email:  sestito@mat.unical.it} }\date{} \maketitle
\abstract{ We show that the extensions of quantum correlations
stemming from a \emph{strict} interpretation of the criterion of
reality of Einstein, Podolsky and Rosen  raise the inadequacy  of
their ideal experiment  for the assignment of simultaneous elements
of reality to two incompatible properties. Then, we suggest a
different physical situation enabling the simultaneous assignment of
objective values of two incompatible observables of a spin particle
by means of measurements of two compatible properties of a second
correlated spin particle. }
\section{Introduction}
In standard quantum theory \cite{vn} the condition characterizing
the simultaneous measurability of more observables is the
commutativity of the corresponding operators; interpretative
questions arise  when incompatible properties are involved: in
general the problem of ascribing simultaneous objective values to
non-commuting observables when them both do not undergo an actual
measurement has no answer in standard quantum theory.\par In the
literature other approaches, aiming to \emph{extend} or
\emph{complete} quantum mechanics, discuss such a question; for
instance, in the context of an extended framework of the operational
approach  \cite{PGL}, \emph{unsharp} (or \emph{fuzzy}) observables
are introduced, represented as positive operator valued measures;
the simultaneous measurability of two unsharp observables is
described by the relation - more general than commutativity- of
\emph{coexistence} asking for the existence of a \emph{joint unsharp
observable} whose statistic contains those of the first two.\par A
different approach is maintained by the hidden variable (HV)
theories \cite{1}-\cite{belinf}, assuming  that each specimen of the
physical system possesses objective values for every observable,
fixed by certain unknown parameters ``which would complete the
information carried out by the quantum states" \cite{BC}, making
possible to determine whether  two incompatible properties are
possessed or not by a specimen of the physical system even when they
are not measured; usually, sometimes implicitly, further assumptions
other than the HV hypothesis are introduced, whose validity can be
questioned \cite{khrennikov}-\cite{GS2010}.\par \vskip0.5pc\noindent
In this work we are concerned with the question of ascribing the
simultaneous values of two non-commuting observables within the
framework of Von Neumann approach. In so doing, we take into account
those observables having only $1$ and $-1$ as possible outcomes,
called \emph{two-value} observables.\par Whenever a functional
relation exists between the outcomes of two commuting observables,
it can be expressed in terms of empirical quantum implications,
defined as follows \cite{N95}: \vskip.5pc\noindent (QI) {\sl Quantum
Implication}.\quad {\it Let $A$ and $P$ be two compatible
observables; in the state $\psi$ the correlation $A\rightarrow P$
holds if and only if in a simultaneous measurement of $A$ and $P$
the occurrence of the outcome 1 for $A$ implies the occurrence of
outcome 1 for $P$.} \vskip0.5pc\noindent \vskip0.5pc\noindent In a
seminal paper of 1935 \cite{2}, Einstein, Podolsky and Rosen (EPR)
describe a physical situation able to attain simultaneous knowledge
about two non-commuting quantities of a system ``on the basis of
measurements made on another system that had previously interacted
with it"; more precisely they infer the simultaneous values of
non-commuting quantities by exploiting correlations between
commuting ones\footnote{They conclude that since quantum mechanics
is unable to describe the simultaneous reality of two non-commuting
quantities it is not a complete theory.}. It is crucial for their
argument the following criterion of reality: \vskip.5pc \noindent
(R) {\sl Criterion of Reality.}\quad{\it If, without in any way
disturbing a system, we can predict with certainty the value of a
physical quantity, then there exists an element of physical reality
corresponding to this physical quantity.} \vskip.7pc\noindent Since
its appearance, two different interpretations of the criterion (R)
are specified; EPR interpreted it as follows: \vskip.5pc\noindent
{\sl Wide Interpretation}. {\it For ascribing reality to $P$ it is
sufficient the ``possibility'' of performing the measurement
of $A$ 
whose outcome would allow for the prediction, with certainty, of the
outcome of a measurement of $P$.} \vskip.5pc\noindent Such a wide
interpretation can be replaced by the narrower following one,
maintained by Bohr \cite{3}: \vskip.5pc\noindent {\sl Strict
Interpretation}. {\it To ascribe reality to $P$ the measurement of
$A$, whose outcome would allow for the prediction, must be actually
performed.} \vskip.5pc
\par In the present work we analyze  implications of adopting
either  interpretations of (R) in connection with the question of
ascribing simultaneous elements of reality to two incompatible
properties; in particular, we prove that the physical situation
described by EPR fails in ascribing reality to two non-commuting
quantities which are both not measured if we adopt the strict
interpretation of (R); then, we suggest a different ideal experiment
enabling such an assignment of objective values.\vskip.5pc Before
describing the logical structure of the work, we point out the
crucial role played by the following principle of locality:
\vskip.7pc\noindent (L) {\sl Principle of Locality}.\quad{\it Let
${\mathcal R}_1$ and ${\mathcal R}_2$ be two space-time regions
which are separated space-like. The reality in ${\mathcal R}_2$ is
unaffected by operations performed in ${\mathcal R}_1$.} \vskip.5pc
\noindent In fact, when considered together, the principle of
locality and the criterion of reality entail an extension of the
validity of quantum correlations. The analysis in \cite{NS2011}
makes evident that when considered with the wide interpretation of
(R), (L) implies the following statement: \vskip.5pc\noindent (EQC)
{\sl Extension of quantum correlations}.\quad {\it Let $A$ and $B$
be two observables whose measurements require operations confined in
two space-time regions ${\mathcal R}_A$ and ${\mathcal R}_B$,
respectively, separated space-like from each other. If quantum
mechanics predicts the perfect correlation $A\rightarrow B$ and
$B\rightarrow A$, in the state $\psi$,  between the outcomes of
actually performed measurements of $A$ and $B$, then every
individual physical system $x$ in the state $\psi$ possesses
objective values of $A$ and $B$ which exhibit the same perfect
correlation.} \vskip0.5pc\noindent A smaller extension, (sEQC),
follows by considering (L) together with the strict interpretation
of (R): \vskip.5pc\noindent (sEQC) {\sl Strict extension of quantum
correlations}.\quad {\it Let $A$ and $B$ be two observables whose
measurements require operations confined in two space-time regions
${\mathcal R}_A$ and ${\mathcal R}_B$, respectively, separated
space-like from each other. If quantum mechanics predicts the
perfect correlation $A\rightarrow B$ and $B\rightarrow A$, in the
state $\psi$, between the outcomes of actually performed
measurements of $A$ and $B$, then every individual physical system
$x$ in the state $\psi$ which actually undergoes a measurement of at
least one within $A$ or $B$ possesses objective values of $A$ and
$B$ which satisfy the same perfect correlation.} \vskip0.5pc
\par
In \cite{NS2011} it has been shown that (EQC) allows  for
simultaneous knowledge of non-commutative observables but conflicts
with locality; on the other hand it cannot be implied by a strict
interpretation of the criterion of reality, so that its validity can
be questioned. On the contrary, (sEQC) is consistent with locality;
does (sEQC) allow for simultaneous knowledge of non-commuting
observables? In other words, can the simultaneous knowledge of
non-commuting observables be consistent with locality? In the
present work we prove that the affirmative answer is valid, by
designing an explicit example. \vskip.5pc The logical structure of
our work is the following: in section 2 we remind the formalism,
extensively introduced in \cite{NS2011}, in order to express
statements (EQC) and (sEQC) in a suitable form. In section 3 we
carry out a formal analysis of the physical situation described by
EPR \cite{2} showing that by replacing (EQC) by (sEQC) this ideal
experiment no longer allows to assign simultaneous element of
reality to two incompatible quantities. In section 4, we propose a
different ideal experiment enabling to do this by means of a spin
measurement of a particle which, in the selected quantum state,
turns out to be correlated with a second separated spin particle. In
the concluding section 5 we outline some insights of these  results
with regard to consistent quantum theory (\cite{GH}-\cite{GO} and
reference therein).
\section{The formalism}
According to standard quantum theory \cite{vn}, every two-value
observable $A$ is represented by a self-adjoint operator $\hat A$ of
the Hilbert space ${\mathcal H}$ associated with the physical
system, with purely discrete spectrum $\sigma(\hat A)=\{1,-1\}$;
every {\it pure} state of the system is represented by a state
vector $\psi\in{\mathcal H}$, with $\Vert\psi\Vert=1$. The
probability of obtaining the outcome 1 by measuring $A$ in the state
$\psi$ is $p_\psi(A,1)=\langle\psi\mid\frac{1}{2}({\bf 1}+\hat
A)\psi\rangle$. \vskip.5pc Given a state vector $\psi$, we define
{\it support} of $\psi$  any concrete non empty set ${\mathcal
S}(\psi)$ of individual physical systems ({\it specimens}) whose
quantum state is $\psi$.
\par
Given a support ${\mathcal S}(\psi)$, in correspondence with a
two-value observable $A$ we introduce the following subsets of
${\mathcal S}(\psi)$:\\ - by ${\bf A}$ we denote the concrete set of
specimens of ${\mathcal S}(\psi)$ which {\it actually} undergo a
measurement of $A$; by ${\bf A}_+$ (resp., ${\bf A}_-$) we denote
the set of specimens of ${\bf A}$ for which the outcome 1 (resp.,
-1) of $A$ has been obtained. Hence, we can assume that ${\bf
A}_+\cup {\bf A}_-={\bf A}$ holds;\\
- by $\tilde{\mathcal A}$ we denote the set of the specimens in
${\mathcal S}(\psi)$ which objectively possess a value of the
observable $A$, without being measured (for instance as a
consequence   of (R)); by $\tilde{\mathcal A}_+$ (resp.,
$\tilde{\mathcal A}_-$) we denote the set of specimens of
$\tilde{\mathcal A}$ which possess the objective value 1 (resp., -1)
of $A$; hence, we can assume that $\tilde{\mathcal A}_+\cup
\tilde{\mathcal A}_-=\tilde{\mathcal A}$ holds.\par We define
${\mathcal A}=\tilde{\mathcal A}\cup{\bf A}$, ${\mathcal
A}_+=\tilde{\mathcal A}_+\cup{\bf A}_+$, ${\mathcal
A}_-=\tilde{\mathcal A}_-\cup{\bf A}_-$. \par Now, we can introduce
the following two mappings $a:{\mathcal A}\to\{1,-1\}$ and ${\bf
a}:{\bf A}\to\{1,-1\}$.
$$a(x)=\left\{
\begin{array}{ll}\;\;\;1,\hbox{ if }x\in{\mathcal A}_+\, ,\\
-1, \hbox{ if } x\in{\mathcal A}_-\, ;\\
 \end{array}\right.
 \quad
{\bf a}(x)=\left\{
\begin{array}{ll}\;\,\;1,\hbox{ if }x\in{\bf A}_+\, ,\\
-1,\hbox{ if }x\in{\bf A}_-\, .\\
 \end{array}\right.
\eqno(1)
$$
The following statements  relate the formalism of standard quantum
mechanics with the physical concepts so far introduced.\vskip.5pc
Given a two-value observable $A$,  since for any $\psi$ we must have
$p_\psi(A,1)\neq 0$ or $p_\psi(A,-1)\neq 0$, we derive the following
statement.
$$
\hbox{If}\quad A\hbox{ is a two-value
observable}\quad\hbox{then}\quad\forall\psi, \;{\mathcal
S}(\psi)\hbox{ exists such that }{\bf A}\neq\emptyset.\eqno(2.o)
$$
According to standard quantum theory,  the following statements can
be assumed to hold for the simultaneous measurability between two
observables $A$ and $B$:
$$
[\hat A, \hat B]\neq{\bf 0}\quad\hbox{implies}\quad {\bf A}\cap{\bf
B}=\emptyset\hbox{ for all }{\mathcal S}(\psi).\eqno(2.i)
$$
$$
[\hat A, \hat B]={\bf 0}\quad\hbox{ implies }
\quad\forall\psi\;\;\exists{\mathcal S}(\psi)\hbox{ such that }{\bf
A}\cap{\bf B}\neq\emptyset.\eqno(2.ii)
$$
Statement (2.ii) merely asserts that $[\hat A,\hat B]={\bf 0}$
ensures the concrete possibility of performing measurement of $A$
and $B$ simultaneously.\par The correlation $A\to B$ in the quantum
state $\psi$ can be formulated in several equivalent ways:
\vskip.5pc $A\to B$ \quad if and only if\quad ${\bf A}_+\cap {\bf
B}\subseteq {\bf B}_+$\quad if and only if ${\bf B}_-\cap {\bf
A}\subseteq {\bf A}_-$, $\forall {\mathcal S}(\psi)$,\par  if and
only if\quad $({\bf a}(x)+1)({\bf b}(x)-1)=0$ for all $x\in{\bf
A}\cap {\bf B}$ whenever ${\bf A}\cap {\bf B}\neq\emptyset$.
\subsection{Extensions of quantum correlations}
The conditions of locality and reality (R,L) lead to further
implications for \emph{separated} observables.  Let $A$ and $B$ be
separated two-value observables, written $A\bowtie B$, i.e.
observables whose measurements require operations confined in
space-like separated regions ${\mathcal R}_A$ and ${\mathcal R}_B$.
As a consequence of the locality condition (L), the following
statement holds.
$$
A\bowtie B\quad\hbox{implies}\quad[\hat A,\hat B]={\bf 0},\;\hbox{
hence } {\mathcal S}(\psi) \hbox{ exists such that } {\bf A}\cap{\bf
B}\neq\emptyset. \eqno(3.i)
$$
\par
Let us suppose that $A\bowtie B$ holds, and  that $A$ is measured on
$x\in{\bf A}$ obtaining ${\bf a}(x)=1$, i.e. $x\in{\bf A}_+$. If the
correlation $A\to B$ also holds, then the prediction of the outcome
1 can be considered valid for a measurement of $B$ on the same
specimen. Now, by (L) the act of actually performing the measurement
of $A$ does not affect the reality in ${\mathcal R}_B$; hence the
criterion (R) could be applied to conclude that $x\in{\mathcal B}$
and $b(x)=1$:
$$ \hbox{if}\quad
A\bowtie B\hbox{ and } A\to B\quad\hbox{then}\quad x\in{\bf
A}_+\Rightarrow x\in{\mathcal B}_+\,.\eqno(3.ii) $$ It is evident
that statement (3.ii) simply follows from the {\sl strict}
interpretation of criterion (R). Analogously, if an actual
measurement of $B$ yields the outcome $-1$, i.e. if $x\in{\bf B}_-$,
then the strict interpretation of (R) leads us to infer that $x\in
{\mathcal A}$ and $a(x)=-1$. Therefore it follows that ${\bf
B}_-\subseteq {\mathcal A}_-\subseteq{\mathcal A}$ and that the
correlation $(a(x)=1)\Rightarrow (b(x)=1)$ also holds for every
$x\in{\bf B}_-$. Hence, according to the strict interpretation of
the criterion (R) the correlation $(a(x)=1)\Rightarrow (b(x)=1)$,
besides holding for all $x\in{\bf A}\cap{\bf B}$, extends to ${\bf
A}_+\cup{\bf B}_-$. Thus, from (R,L) and quantum mechanics we infer
the following statement. \vskip.5pc\noindent (sR) {\it Let $A$ and
$B$ be space-like separated two-value observables. If $A\to B$ then
$$\quad(a(x)+1)(b(x)-1)=0,\;\;\forall x\in({\bf
A}_+\cup{\bf B}_-)\cup({\bf A}\cap{\bf B}) .\eqno(4.i)
$$}
The quantum correlation $A\leftrightarrow B$, i.e. $A\to B$ and
$B\to A$, in the state $\psi$ means that  the correlation
$(a(x)=1)\Leftrightarrow (b(x)=1)$ holds for all $x\in{\bf
A}\cap{\bf B}$ for all ${\mathcal S}(\psi)$. In this case, from
(sEQC) we can deduce that $(a(x)=1)\Leftrightarrow (b(x)=1)$ holds
for all $x\in({\bf A}_+\cup{\bf B}_-)\cup({\bf B}_+\cup{\bf
A}_-)\cup({\bf A}\cap{\bf B})={\bf A}\cup{\bf B}$ for all ${\mathcal
S}(\psi)$. Hence, (sR) incorporates the strict extension (sEQC) of
quantum correlation $A\leftrightarrow B$ in the state $\psi$:
$$A\bowtie B,\;A\leftrightarrow B\quad\hbox{imply}
\quad {\bf A}\cup{\bf B}\subseteq{\mathcal A}\cap{\mathcal
B}\;\hbox{i.e. } a(x)=b(x),\; \forall x\in{\bf A}\cup{\bf
B},\;\forall{\mathcal S}(\psi). \eqno(4.ii)
$$
\vskip.5pc The wide interpretation of criterion (R) allows for
larger extensions. Indeed it leads us to infer the following
statements:
$$
\hbox{If}\quad A\bowtie B\hbox{ and } A\to B\quad\hbox{then}\quad
{\mathcal A}_+\subseteq{\mathcal B}_+\hbox{ and } {\mathcal
B}_-\subseteq{\mathcal A}_-,\, \forall{\mathcal
S}(\psi);\eqno(5.i)$$
$$
\hbox{If}\quad A\bowtie B\hbox{ and } A\leftrightarrow
B\quad\hbox{then}\quad {\mathcal A}_+={\mathcal B}_+\hbox{ , }
{\mathcal B}_-={\mathcal A}_-\hbox{ and }{\mathcal A}={\mathcal
S}(\psi),\;\forall{\mathcal S}(\psi).\eqno(5.ii)$$ The statement
(5.ii) is nothing else but (EQC) stated in formal terms. The
statement (5.i) says that the correlation ``$a(x)=1$ implies
$b(x)=1$'' extends to ${\mathcal A}_+\cup {\mathcal B}_-$.
\section{The criterion of reality and the experiment of Einstein, Podolsky and Rosen}
In this section we show that while EPR's argument can be used to
infer simultaneous elements of reality for non-commuting properties
when (EQC) is adopted, this no longer holds if (sEQC) replaces
(EQC). In so doing, we  recur to the simplified form of EPR
experiment suggested in \cite{Bhom}. \vskip.5pc The system is made
up of a pair of separated non interacting spin-$1/2$ particles, in
the singlet state $\psi$. Let us denote the two spin observables of
the first (resp., second) particle along two fixed non parallel
directions by $A$ and $B$ (resp., $P$ and $Q$); in
$\frac{1}{2}\hbar$ units, they are two-value observables. According
to quantum mechanics, if we actually measure a spin component, $A$
or $B$,  of the first particle then the outcome of an actual
measurement of the same component for the second particle, $P$ or
$Q$ respectively, turns out to be the opposite. Hence, in the state
$\psi$, the correlations $A\leftrightarrow -P$ and $B\leftrightarrow
-Q$ hold, i.e. for any ${\mathcal S}(\psi)$ the following statements
hold for concrete outcomes: {\setlength\arraycolsep{2pt}
$$
\begin{array}{lll}
\textrm{   i)}\quad &{\bf a}(x)={\bf -p}(x)\quad &\forall x\in({\bf
A}\cap{\bf P}),\\ \textrm{ ii)}\quad &{\bf b}(y)={\bf -q}(y)\quad
&\forall y\in({\bf B}\cap{\bf Q}).\\
\end{array}
\eqno(6.i)
$$}
In \cite{2}, by means of the criterion of reality, EPR provide the
following argument entailing an extension of the validity of such
correlations. \vskip.5pc\noindent {\sl EPR's argument}. By measuring
either $A$ or $B$ we can predict with certainty, and without in any
way disturbing the system, either the value of $P$ or the value of
$Q$; so, according to (R), in the first case $P$ is an element of
reality, in second one $Q$ is an element of reality, arriving at the
conclusion that two physical incompatible observables have
simultaneous reality. \vskip.5pc\noindent Such a statement is
noticeably supported by the wide interpretation of (R); in fact,
since $A$ and $B$ are non-commuting quantities, they cannot be
measured together. As a consequence (EQC) holds, then correlations
(6.i) can be extended to the following correlations between
objective values.{\setlength\arraycolsep{2pt}
$$
\left.
\begin{array}{llll}
\textrm{   i)}\quad &{ a}(x)=-{ p}(x),\\
\textrm{  ii)}\quad &{ b}(x)=-{ q}(x).\\
\end{array}\right\}
\quad\forall x\in{\mathcal S}(\psi).
$$}
\noindent Hence, in spite of the incompatibility between $A$ and $B$
and the consequent  impossibility of measuring them together, every
specimen $x\in{\mathcal S}(\psi)$ possesses  values satisfying the
correlations predicted by quantum mechanics. \vskip.5pc However, the
validity of (EQC) is questioned in \cite{NS2011} where it is shown
to be responsible for the inconsistency between quantum mechanics
and locality claimed by the non-locality theorems  of Hardy
\cite{4}, of Greenberger, Horne, Shimony and Zeilinger \cite{5} and
of Bell \cite{1}; moreover, the quoted inconsistency is proved to
disappear if (sEQC) replaces (EQC).\par For this reason, we
investigate the consequences of adopting (sEQC), instead of (EQC),
in connection with EPR's argument. In particular, we prove that the
inferred simultaneous reality of both $P$ and $Q$ no longer holds if
the strict interpretation of (R) is adopted. In such a case,  the
extensions of correlations (6.i) are obtained by applying
(4.i):{\setlength\arraycolsep{2pt}
$$
\begin{array}{lll} \textrm{ i)} \quad &a(x)=-p(x),
\quad &\forall x \in\mathbf{A}\cup \mathbf{P}\equiv \mathbf{X}\\
\textrm{ii)} \quad &b(y)=-q(y), \quad &\forall y \in\mathbf{B}\cup
\mathbf{Q}\equiv \mathbf{Y}.
\end{array}
\eqno(7)$$ In order to ascribe simultaneous reality to $P$ and $Q$,
(7.i) and (7.ii) should hold for the same specimen $x_0\in
\mathbf{X}\cap \mathbf{Y}$. From (2.i) we derive
$$\begin{array}{ll}
\mathbf{X}\cap \mathbf{Y}&= (\mathbf{A}\cup \mathbf{P})\cap
(\mathbf{B}\cup \mathbf{Q})=\\
&=(\mathbf{A}\cap \mathbf{B})\cup(\mathbf{A}\cap \mathbf{Q})\cup
(\mathbf{P}\cap \mathbf{B})\cup (\mathbf{P}\cap \mathbf{Q})=\\
&=(\mathbf{A}\cap \mathbf{Q})\cup (\mathbf{P}\cap \mathbf{B}).
\end{array}$$}
Since $\mathbf{X}\cap \mathbf{Y}\neq \emptyset$, by (2.ii) one could
think that some specimen $x_0\in {\mathcal S}(\psi)$ exists,
satisfying all the requirements. However, if $x_0\in \mathbf{P}\cap
\mathbf{B}$  we are not allowed to ascribe simultaneous reality to
$A$ and $Q$; indeed, the principle of locality ensures that the
reality in ${\cal R}_2$ is not affected by operations performed in
${\cal R}_1$ but an actual measurement of $P$ occurs just in the
regions ${\cal R}_2$; for instance, let us suppose that   $x_0\in
\mathbf{P}\cap \mathbf{B}$; since $x_0\in \mathbf{B}$,  locality
cannot be invoked for deducing that $a(x_0)=-p(x_0)$ because the
measurement of $\mathbf{B}$ could affect the value of $A$.
\par \vskip1pc We conclude that, by replacing
the wide interpretation of the criterion of reality with the strict
one, the example of EPR does not allow to  ascribe reality to two
non commuting quantities.
\section{An ideal experiment} In this section we describe  an
ideal experiment enabling to ascribe simultaneous reality to two
incompatible observables when the strict interpretation of the
criterion of reality is adopted.\par The observables involved are
\emph{0-1 observables}, i.e. having 0 and 1 as possible outcomes,
represented in the theory by projection operators. In such a case
statements (4.ii) and (5.ii), involved in our argument, while
derived for two-value observables, turn out to be valid for 0-1
observables.\par As in the previous ideal experiment we exploit two
quantum correlations of the type $A\leftrightarrow P$, for which
(sEQC) implies the extension (4.ii); hence, whenever a measurement
of $P$ is actually performed, we can consider ``objective'' the
observable $A$, i.e. for every specimen of the physical system which
undergoes the measurement of $P$ we are able to infer whether $A$ is
possessed or not possessed by the specimen.\par As a consequence,
the physical situation described in the rest of this section not
only entails the simultaneous reality of two incompatible
properties, but also provides their objective values.
 \vskip.5pc The physical system consists of two separated and non-interacting
particles, $I$ and $II$. Particle $I$  is a spin-5/2 particle
localized in a region ${\mathcal R}_I$  and described in the Hilbert
space ${\cal H}_I$; particle $II$ is a spin-3/2 particle localized
in a region ${\mathcal R}_{II}$ and described in the Hilbert space
${\cal H}_{II}$; therefore, ${\cal H}_I\otimes {\cal H}_{II}$ is the
Hilbert space describing the entire system. We adopt the
Heisenberg's picture; in the notation for operators, the suffix $I$
(resp. $II$) denotes an operator of ${\cal H}_I$ (resp., ${\cal
H}_{II}$). By $A_{II}^1$ we denote the projection operator of ${\cal
H}_{II}$ representing the event ``the spin component of particle
$II$ in the z-direction is $3/2$"; similarly, we define the
projections $A_{II}^2$, $A_{II}^3$, $A_{II}^4$ associated to the
values $1/2$, $-1/2$, $-3/2$ of the spin along $z$, respectively. We
denote their respective eigenvectors relative to the eigenvalue 1 by
$|\frac{3}{2}\rangle_{II}$, $|\frac{1}{2}\rangle_{II}$,
$|-\frac{1}{2}\rangle_{II}$, $|-\frac{3}{2}\rangle_{II}$. By
$A_{I}^1$ we denote the projection operator of ${\cal H}_{I}$
representing the event ``the spin component in the z-direction is
$5/2$"; similarly, we define the projections $A_{I}^2$, $\dots$,
$A_{I}^6$ associated to the values $3/2$, $1/2$, $-1/2$, $-3/2$,
$-5/2$ of the spin along $z$, respectively. We denote their
respective eigenvectors relative to the eigenvalue 1 by
$|\frac{5}{2}\rangle_{I}$, $|\frac{3}{2}\rangle_{I}$,
$|\frac{1}{2}\rangle_{I}$, $|-\frac{1}{2}\rangle_{I}$,
$|-\frac{3}{2}\rangle_{I}$, $|-\frac{5}{2}\rangle_{I}$. \par Let us
now introduce three projection operators $B_I^i$, with $i=1,2,3$,
where $B_I^i=|\psi_1^i\rangle\langle \psi_1^i|$ and
$|\psi_1^1\rangle=\frac{1}{2}(|\frac{5}{2}\rangle_I-
|\frac{3}{2}\rangle_I+ |-\frac{1}{2}\rangle_I-
|-\frac{3}{2}\rangle_I)$, $|\psi_1^2\rangle=|\frac{1}{2}\rangle_I$
and $|\psi_1^3\rangle=|-\frac{5}{2}\rangle_I$.\\
One of two non-commuting observables, $E$ or $G$, can be measured on
system $I$, where:
$$
\begin{array}{ll}
E=E_{I}\otimes \textbf{1}_{II}=(A_I^1+A_I^2+A_I^3)\otimes
\textbf{1}_{II}\\
G=G_{I}\otimes \textbf{1}_{II}=(B_I^1+B_I^2+ B_I^3)\otimes
\textbf{1}_{II}
\end{array}
\eqno(8)$$ Now we consider the projection operators
$T=\textbf{1}_I\otimes (A_{II}^1+ A_{II}^2)$ and
$Y=\textbf{1}_I\otimes (A_{II}^1+ A_{II}^3)$; straightforward
calculations show that $T$ and $Y$ represent commuting properties of
particle $II$ so that a support ${\mathcal S}_0(\psi)$ exists such
that $\mathbf{T}\cap \mathbf{Y}\neq\emptyset$; furthermore, their
measurements require operations confined in the region ${\mathcal
R}_{II}$, so that they are separated from, hence commuting (with),
both $E$ and $G$.\par Let the system be prepared in the entangled
state represented by\vskip.5pc\noindent $\psi=\frac{\sqrt
3}{4}(|\frac{5}{2}\rangle_I+
|\frac{3}{2}\rangle_I)|\frac{1}{2}\rangle_{II}+ \frac{1}{\sqrt
8}|\frac{1}{2}\rangle_I|\frac{3}{2}\rangle_{II}+
\frac{1}{4}(|-\frac{1}{2}\rangle_I+|-\frac{3}{2}\rangle_I)\mid-\frac{3}{2}\rangle_{II}+
\sqrt{\frac{3}{8}}
\mid-\frac{5}{2}\rangle_I\mid-\frac{1}{2}\rangle_{II}.$\vskip.5pc In
the state $\psi$, projection operators $T$ and $E$ turn out to
satisfy the condition $E\psi=T\psi$, which is equivalent to the
following relation involving conditional probabilities:
$$p(E|T)=\frac{\langle \psi|ET\psi\rangle}{\langle
\psi|T\psi\rangle}=1=\frac{\langle \psi|TE\psi\rangle}{\langle
\psi|E\psi\rangle}=p(T|E).\eqno(9)$$
According to quantum mechanics this is equivalent to say that in a
simultaneous measurement of $T$ and $E$, outcome 1 (resp., 0) for
$T$ (resp., $E$) ensures outcome 1 (resp., 0) for $E$ (resp., $T$),
i.e. $E\leftrightarrow T$ and equivalently
$$e(x)=t(x)\quad \forall x\in {\mathbf E}\cap{\mathbf T}, \quad\forall {\mathcal S}(\psi).$$
Similarly, equation $G\psi=Y\psi$ holds, entailing $p(G|Y)=p(Y|G)=1$
for the conditional probabilities, and equivalent to the quantum
correlation $G\leftrightarrow Y$, i.e.
$$g(z)=y(z)\quad \forall z\in {\mathbf G}\cap{\mathbf Y}, \quad\forall {\mathcal S}(\psi).$$
\vskip.5pc Now we prove that if we adopt (sEQC), the envisaged
physical situation allows to  ascribe the simultaneous objective
values to  two incompatible properties.\par Statement (sEQC)
incorporates the following extensions of the quantum correlations in
the state $\psi$: $$\begin{array}{ll} i)\quad & e(x)=t(x)\quad
\forall x\in {\mathbf E}\cup{\mathbf T}={\mathbf X}, \quad\forall
{\mathcal S}(\psi);\cr ii) \quad & g(z)=y(z)\quad \forall z\in
{\mathbf G}\cup{\mathbf Y}={\mathbf Z}, \quad\forall {\mathcal
S}(\psi).\end{array} \eqno(10)$$ In the state $\psi$, for any $x\in
\mathbf{T}\cap \mathbf{Y}\subseteq \mathbf{X}\cap \mathbf{Y}$
extensions (10.i) and (10.ii) hold, so that we can conclude that
from the outcomes of actual measurements of $T$ and $Y$ we can infer
both the objective values of $E$ and of $G$, in spite of their
incompatibility, according to the following table. \vskip10pt
\centerline{\begin{tabular}{|c|c|c|c|}
\hline \quad \bf{T}\quad  &\quad \bf{Y}\quad &\quad E\quad & G\quad\\
\hline\quad 1\quad &\quad 1\quad &\quad 1\quad &\quad 1\quad\\
\hline\quad 1\quad &\quad 0\quad &\quad 1\quad &\quad 0\quad\\
\hline\quad 0\quad &\quad 1\quad &\quad 0\quad &\quad 1\quad\\
\hline\quad 0\quad &\quad 0\quad &\quad 0\quad &\quad 0\quad\\
\hline
\end{tabular}}
\par\vskip1pc
Notice that in the present ideal experiment, the act of ascertaining
the value of $E$ does not affect the value of the other observable
$G$, contrary to what happens in EPR experiment, because the
measurement of $T$ and $Y$ are performed in region ${\mathcal
R}_{II}$, which is space-like separated from ${\mathcal R}_{I}$. 
\section{Simultaneous reality of incompatible properties in the consistent quantum theory}
The assignment of reality to the non-commuting properties $E$ and
$G$ of section 4 makes them objective, though not measured. Such a
result  has interesting insights in the context of the consistent
quantum theory (CQT) (\cite{GH} and references therein). CQT is an
extension of standard quantum theory where the basic concept of
\emph{event} (of standard quantum theory) is generalized to that of
\emph{history}, defined as a finite sequence $h=(E_1,E_2,\ldots,
E_n)$ of events that the system objectively possesses at respective
times $t_1$, $t_2$, $\ldots$, $t_n$. CQT establishes that when a
family of histories ${\mathcal C}$ satisfies a criterion of
\emph{consistency} then \vskip.5pc \noindent (I) {\it the set of all
``elementary" histories of ${\mathcal C}$ is a ``sample space of
mutually exclusive elementary events, one and only one of which
occurs"} \cite{GH}. \vskip.7pc\noindent The \emph{occurrence} of a
history has to be interpreted as follows: \vskip.5pc \noindent (O)
{\it A given history $h=(E_1,E_2,\ldots,E_n)$ occurs if all events
$E_1$, $E_2$, $\ldots$, $E_n$ objectively occur at respective times
$t_1$, $t_2$, $\ldots$, $t_n$. The occurrence of a history is an
objective fact, independent of the performance of a measurement that
reveals this occurrence.} \vskip.7pc\par The criterion of
consistency postulated by CQT is the following: \vskip.5pc \noindent
(C) {\it A family ${\cal C}$ is consistent, in the sense of
definition above, if and only if it is weakly decohering. i.e.
condition $Re(Tr(C_{h_1}\rho C^{\ast}_{h_2}))=0$ holds for all
mutually exclusive histories $h_1$ and $h_2$, where $C_h=E_n\cdot
E_{n-1}\cdots E_1$. In this case $p(h)= \frac{1}{N}Tr(C_{h}\rho
C^{\ast}_{h})$ is the probability of occurrence history $h$.}
\vskip.7pc\par According to CQT, conclusions drawn in two different
families, ${\cal C}_1$ and ${\cal C}_2$, hold together in the case
that these families are \emph{compatible}, i.e. if a third
consistent family ${\cal C}$ exists  such that ${\cal C}_1\cup {\cal
C}_2\subseteq {\cal C}$.
\par Moreover, let $h=(E_1, {\bf 1})$ a two-time history and let
${\cal C}(h)$ be the smallest family containing $h$; the possibility
exists of establishing if a system possesses property $E_1$ by means
of the measurement of a different observable $E_2$ at time $t_2$.
Indeed, history $h_1=(E_1, E_2)$ is a refinement of $h$; hence,
${\cal C}(h_1)$ is a refinement of ${\cal C}(h)$, i.e. ${\cal
C}(h)\subseteq {\cal C}(h_1)$; if standard quantum theory predicts
that $p(E_1|E_2)=1$ then a measurement of $E_2$ with concrete
outcome 1, together with (O), reveals that $h$ occurred. \vskip.7pc
The analysis of the conceptual problems raised by CQT led some
authors to extend the conceptual basis of CQT \cite{nis2005}. Then,
for every family ${\cal C}$ the existence of a \emph{support} of
${\mathcal C}$ is postulated, defined  as the concrete set
$b({\mathcal C})$ of all specimens of the physical system such that
for each individual $s\in b({\mathcal C}) $ every history of
${\mathcal C}$ either occurs or does not occur (briefly, \emph{makes
sense}).  Accordingly, a family ${\mathcal C}$ is \emph{consistent}
if and only if $b({\mathcal C})\neq \emptyset$. Given a history $h$
in such a family, by $b_1(h)$ (resp., $b_0(h)$) we denote the subset
of those systems for which   $h$ occurs (resp., does not occur).\par
The concept of incompatible families of standard CQT within the
formalism of the extended basis becomes:
\begin{itemize}
\item[i)] Let ${\cal C}_1$ and ${\cal C}_2$ be two incompatible
families, then $b({\cal C}_1)\cap b({\cal C}_2)=\emptyset$.
\end{itemize}
The possibility claimed in CQT of revealing the occurrence of $E$ by
means of the outcome of a measurement of $T$ can be expressed by the
following \emph{condition of objectification}:
\begin{itemize}
\item[ii)] for all $s\in b({\cal C}(h_E))$, $s\in b_1(h_T)$
implies $s\in b_1(h_E)$.
\end{itemize}
Moreover, the following statement is assumed to hold in the extended
conceptual basis of CQT \cite{nis2005}:
\begin{itemize}
\item[iii)]
Let ${\cal C}_1$ and ${\cal C}_2$ be two families of histories; then
${\cal C}_1\subseteq {\cal C}_2$ implies $b({\cal C}_2)\subseteq
b({\cal C}_1)$.
\end{itemize}
\vskip.7pc In the rest of this section we analyze consequences of
adopting the extensions of correlations (sEQC) in connection with
CQT. \vskip.7pc In the previous section we designed a physical
situation making objective two incompatible properties, $E$ and $G$,
by means of measurements of (compatible) $T$ and $Y$. We can
consider the events $E$ and $G$ at a time $t_1$ immediately prior
the events $T$ and $Y$ at a time $t_2$; in so doing, two consistent
families of histories naturally arise: the families ${\cal C}(h_E)$
and ${\cal C}(h_G)$ where $h_E=(E,T)$ and $h_G=(G,Y)$. They are
refinements of families ${\cal C}(h_T)$ and ${\cal C}(h_Y)$
respectively, generated by histories $h_T=({\bf 1}, T)$ and
$h_Y=({\bf 1}, Y)$; so that (iii) implies $b({\cal C}(h_E))\subseteq
b({\cal C}(h_T))$ and $b({\cal C}(h_G))\subseteq b({\cal C}(h_Y))$.
As a consequence, in general, for a specimen $s\in b_1(h_T)$,
condition (ii) does not entail that $s\in b_1(h_E)$, although in the
state $\psi$ the correlation $E\leftrightarrow T$ holds, unless
$s\in b({\cal C}(h_E))$.
\par Let us suppose that for a specimen $s$ both $T$ and
$Y$ occur, i.e. $s\in b_1(h_T)\cap b_1(h_Y)$; since  we defined
$T=A_1+A_2$ and $Y=A_1+A_3$, where $A_i\bot A_j$ for $ i\neq j$,
$i=1,2,3,4$, and $\sum_i A_i={\bf 1}$, (I) implies that the
elementary history $h_{A_1}=({\bf 1}, A_1)$ occurs for $s$ hence,
$s\in b_1(h_{A_1})$.
\par
Standard quantum theory predicts a non-vanishing probability of
occurrence for $A_1$, $p(A_1)\neq 0$; as a consequence,
$b_1(h_{A_1})\neq\emptyset$ so that $s_0$ exists such that $s_0\in
b_1(h_{A_1})$; furthermore, (iii) implies $s_0\in b_1(h_{T})\cap
b_1(h_{Y})$.\par In the state $\psi$, the extension of quantum
correlations (10) implies that for any $s_0\in b_1(h_{T})\cap
b_1(h_{Y})$ we can infer both $s_0\in b_1(h_{E})$ and $s_0\in
b_1(h_{G})$. Then, we have to conclude that
$$s_0\in b({\cal
C}(h_{E}))\cap b({\cal C}(h_{G}))\neq 0.\eqno(11) $$ But $[E,G]\neq
0$, hence the families ${\cal C}(h_{E})$  and ${\cal C}(h_{G})$ are
incompatible; as a consequence (11) contradicts (i). Thus, the
possibility of a double assignment of objective values two
non-commuting properties, provided in the present work, gives rise
to interesting interpretative question in connection with CQT,
opening the possibility that an individual specimen $s$ of the
physical system can simultaneously follow histories $h_E\in {\cal
C}_E$ and $h_G\in {\cal C}_G$,  though no consistent family exists
containing both of them.


\end{document}